\documentclass[conference]{IEEEtran}
\IEEEoverridecommandlockouts
\usepackage{cite}
\usepackage{amsmath,amssymb,amsfonts}
\usepackage{algorithmic}
\usepackage{graphicx}
\usepackage{textcomp}
\usepackage{xcolor}
\def\BibTeX{{\rm B\kern-.05em{\sc i\kern-.025em b}\kern-.08em
    T\kern-.1667em\lower.7ex\hbox{E}\kern-.125emX}}

\usepackage{url}

\usepackage{hyperref}
\usepackage{listings}
\definecolor{eclipseStrings}{RGB}{42,0.0,255}
\definecolor{eclipseKeywords}{RGB}{127,0,85}
\colorlet{punct}{red!60!black}
\definecolor{delim}{RGB}{20,105,176}
\colorlet{numb}{black}

\definecolor{background}{HTML}{EEEEEE}

\lstdefinelanguage{json}{
    basicstyle=\normalfont\ttfamily,
    commentstyle=\color{eclipseStrings},
    stringstyle=\color{eclipseKeywords},
    numberstyle=\scriptsize,
    numbersep=8pt,
    showstringspaces=false,
    breaklines=true,
    frame=lines,
    backgroundcolor=\color{background},
    string=[s]{"}{"},
    comment=[l]{:\ "},
    morecomment=[l]{:"},
    literate=
     *{0}{{{\color{numb}0}}}{1}
      {1}{{{\color{numb}1}}}{1}
      {2}{{{\color{numb}2}}}{1}
      {3}{{{\color{numb}3}}}{1}
      {4}{{{\color{numb}4}}}{1}
      {5}{{{\color{numb}5}}}{1}
      {6}{{{\color{numb}6}}}{1}
      {7}{{{\color{numb}7}}}{1}
      {8}{{{\color{numb}8}}}{1}
      {9}{{{\color{numb}9}}}{1}
      {:}{{{\color{punct}{:}}}}{1}
      {,}{{{\color{punct}{,}}}}{1}
      {\{}{{{\color{delim}{\{}}}}{1}
      {\}}{{{\color{delim}{\}}}}}{1}
      {[}{{{\color{delim}{[}}}}{1}
      {]}{{{\color{delim}{]}}}}{1},
}

\lstdefinelanguage{http}{
    breaklines=true,
    frame=lines,
    backgroundcolor=\color{background},
}

\newif\ifll
\newcommand{\longversion}[1]{\ifll{#1}\fi}
\llfalse

\begin{document}

\title{Smart Bulbs can be Hacked to Hack into your Household*\\
\thanks{In Proceedings of the 20th International Conference on Security and Cryptography (\href{https://secrypt.scitevents.org/Home.aspx?y=2023}{Secrypt 2023}), ISBN 978-989-758-666-8, ISSN 2184-7711, pages 218-229.  DOI: \href{https://doi.org/10.5220/0012092900003555}{10.5220/0012092900003555}
}
}

\author{\IEEEauthorblockN{1\textsuperscript{st} Davide Bonaventura}
\IEEEauthorblockA{\textit{Dipartimento di Matematica e Informatica} \\
\textit{Universit\`a di Catania}\\
Catania, Italy \\
d.bonaventura@studium.unict.it\\ 
\href{https://orcid.org/0009-0004-4463-7991}{orcid.org/0009-0004-4463-7991}}
\and
\IEEEauthorblockN{2\textsuperscript{nd} Sergio Esposito}
\IEEEauthorblockA{\textit{Information Security Group}\\
\textit{Royal Holloway} \\
\textit{University of London}\\
Egham, UK \\
sergio.esposito.2019@live.rhul.ac.uk\\
\href{https://orcid.org/0000-0001-9904-9821}{orcid.org/0000-0001-9904-9821}}
\and
\IEEEauthorblockN{3\textsuperscript{rd} Giampaolo Bella}
\IEEEauthorblockA{\textit{Dipartimento di Matematica e Informatica} \\
\textit{Universit\`a di Catania}\\
Catania, Italy \\
giamp@dmi.unict.it\\ 
\href{https://orcid.org/0000-0002-7615-8643}{orcid.org/0000-0002-7615-8643}}
}

\maketitle

\begin{abstract} 
The IoT is getting more and more pervasive. Even the simplest devices, such as a light bulb or an electrical plug, are made ``smart'' and controllable by our smartphone. 
This paper describes the findings obtained by applying the PETIoT kill chain to conduct a Vulnerability Assessment and Penetration Testing session on a smart bulb, the {Tapo L530E} by {Tp-Link}, currently best seller on Amazon Italy. We found that four vulnerabilities affect the bulb, two of High severity and two of Medium severity according to the CVSS v3.1 scoring system. In short, authentication is not well accounted for and confidentiality is insufficiently achieved by the implemented cryptographic measures. In consequence, an attacker who is nearby the bulb can operate at will not just the bulb but all devices of the Tapo family that the user may have on her Tapo account. Moreover, the attacker can learn the victim's Wi-Fi password, thereby escalating his malicious potential considerably. The paper terminates with an outline of possible fixes.

\end{abstract}


\section{INTRODUCTION}\label{sec:introduction}
The Internet of Things (IoT) surrounds people virtually everywhere due to the increasing number of devices that become computerised and interconnected --- it may even pervade people's bodies thanks to the rise of implantable medical devices and wearable devices. IoT devices exceeded 13.8 billion in 2021 and are expected to double by 2025~\cite{statistics}.
This world of devices leads to a massive attack surface with a significant increase in the number of entry points for hackers and, correspondingly, security and privacy challenges for researchers and engineers to face.

While the full range and inherent diversity of IoT devices cannot be overestimated, safety critical devices, such as self-driving cars, Industrial Control Systems (ICS) and their security and privacy challenges, risk drawing too much attention to the detriment of inexpensive devices, such as those for home automation. It must be stressed that any home automation issue would be highly impactful due to the enormous use of such devices worldwide, which is also favoured by a general price decrease. 

We observe that similar considerations apply to smart bulbs, whose hijacking may have security implications, e.g., help a thief abuse the victim's electricity consumption, then privacy implications, e.g., help the thief profile the victim's patterns of usage hence their habits and, ultimately, safety implications, e.g., ultimately help the thief understand if a household is currently empty or not. Following these observations, this paper rests on the following research question: \textit{What vulnerabilities affect best-seller light bulbs? What consequences do they have on the end user's security, privacy and safety?}

\subsection{Contributions}
\begin{figure}[htbp]
  \centering
   {\includegraphics[scale=0.3]{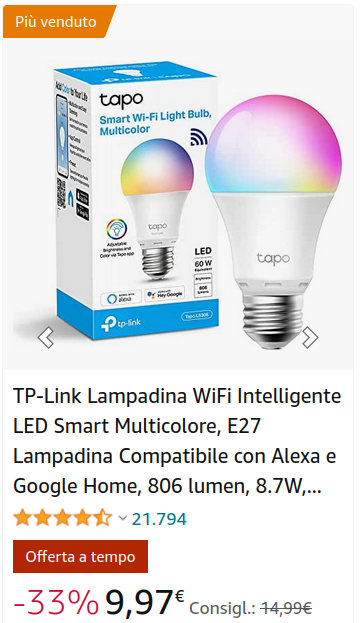}}
  \caption{The Tapo L530E on Amazon.it}
  \label{fig:Tapo_on_amazon}
 \end{figure}

Our experiments apply the (steps of the) PETIoT kill chain~\cite{petiot} to conduct a Vulnerability Assessment and Penetration Testing (VAPT) on the Tp-Link Tapo Smart Wi-Fi Light Bulb, Multicolor (L530E), currently best-seller on Amazon Italy, Fig.~\ref{fig:Tapo_on_amazon}.
The Tapo L530E is a cloud-enabled multicolor Smart Bulb that can be controlled through {the {Tapo} proprietary application}. The user needs to install it on an Android or iOS mobile device and then create a {Tapo} account. The smart bulb uses Wi-Fi technology for connectivity. So, unlike many other smart bulbs requiring a dedicated hub, the user can enjoy the L530E as is by simply connecting it to their home Wi-Fi network. 

Contrarily to a potential belief that smart bulbs are not worth protecting or hacking, we found out that this model suffers four vulnerabilities that are not trivial and, most importantly, may have a dramatic impact:

\begin{enumerate}
    \item Vulnerability 1. \textit{Lack of authentication of the smart bulb with the Tapo app}, 8.8 CVSS score, High severity. The app does not get any guarantee about the identity of its peer. Therefore, anyone can authenticate to the app and pretend to be the smart bulb.

    \item Vulnerability 2. \textit{Hard-coded, short shared secret}, 7.6 CVSS score, High severity. The secret used by both the Tapo app and the smart bulb is short and exposed by both the code fragments run by the app and by the smart bulb.
    
    \item  Vulnerability 3. \textit{Lack of randomness during symmetric encryption}, 4.6 CVSS score, Medium severity. The initialisation vectors (IVs) used by the Tapo app and the smart bulb are static, and each communication session uses a single, fixed IV for each message.

    \item  Vulnerability 4. \textit{Insufficient message freshness}, 5.7 CVSS score, Medium severity. Neither the app nor the smart bulb implements appropriate measures to check the freshness of messages that they receive.
\end{enumerate}
Our exploitation experiments of such vulnerabilities demonstrate that a malicious attacker who stands in proximity of the target smart bulb, hence of the Wi-Fi access point to which the bulb is meant to be connected, can exploit the bulb in various ways. 
Vulnerability 1 means that the attacker impersonates the bulb and receives the user's Tapo credentials as well as the user's Wi-Fi credentials from the Tapo app. To achieve this, the bulb must be in setup mode, when it exposes its own SSID. Alternatively, if the bulb is already configured and working, then the attacker mounts a simple Wi-Fi deauthentication attack against the bulb and repeats it until the user attempts to setup the bulb again to restore it. 
\longversion{In either case, the attacker exposes an SSID under his control mimicking the bulb's SSID and waits for the user to connect to the attacker's one to complete the setup. Should the user choose the genuine SSID, then the attacker would reiterate deauthentication. 
It is while the user thinks she is authenticating with the bulb, in fact impersonated by the attacker, that the Tapo app discloses the user's Tapo credentials. 
In consequence, the attacker may now impersonate the user with any other smart devices of the Tapo family that the user may have, and operate them at will, also with consequences on user safety when the target device, for example, is a smart plug supplying an oven appliance.}

The attacker may also interleave another session: by leveraging the credentials just obtained, he impersonates the user through the setup of the bulb and receives a session key from the device, which he may then relay back to the user. Therefore, the attacker effectively mounts a man-in-the-middle attack. 
Moreover, during device setup, the Tapo app also releases the Wi-Fi credentials to the attacker, thereby causing a clear escalation of the malicious potential for other attacks requiring local access.

Vulnerability 2 means that the attacker can obtain the key that the Tapo app and the smart bulb share and use for the authentication and the integrity of the messages exchanged during the initial discovery phase, described in Section~\ref{sec:bulb-discovery}. Thanks to the knowledge of the key, the attacker can violate the integrity and authentication of the messages of this phase.

Vulnerability 3 means that the attacker understands the consequences of certain encrypted messages on the target device despite the fact that he does not understand the precise cleartext inside each message. Therefore, the attacker may attempt to re-use those messages at will to operate the device as each message determines. In combination with vulnerability 4, the attacker is assured that whatever message he replays will be accepted by the bulb, hence a Denial-of-Service (DoS) attack becomes possible.


\subsection{Ethics and Responsible Disclosure}
All experiments performed on Tapo L530E only involve devices, Wi-Fi networks, accounts, emails, and passwords owned solely by the authors of this work. During the experiments, no user nor third-party data were accessed.

We dutifully contacted Tp-Link via their Vulnerability Research Program (VRP), reporting all four vulnerabilities we found.  
They acknowledged all of them and informed us that they started working on fixes both at the app and at the bulb firmware levels, planning to release them in due course.

\subsection{Paper Summary} 
This paper continues with a short account of the related work (\S\ref{sec:rw}). It then unfolds the six phases of the chosen methodology, i.e., the PETIoT kill chain, on the chosen target, i.e., the Tapo L530E (\S\ref{sec:1} through \S\ref{sec:6}). Finally, it draws the relevant conclusions (\S\ref{sec:concl}).

\section{RELATED WORK}\label{sec:rw}

The security and privacy aspects of IoT devices are becoming more and more important and, correspondingly, the related work concerning such devices is very wide. Due to space constraints, this section is reduced to the literature entries that are most relevant to the core of this paper.

In 2021, it was shown that printers are common devices whose networked use may suffer at least three attacks~\cite{strive19}. The first attack, \textit{zombies for traditional DDoS}, shows how some printers may suffer from vulnerabilities that would transform them into exploitable zombies. The second attack, \textit{paper DoS}, shows how a large number of printers are found to honour unauthenticated printing requests. The third attack, \textit{privacy infringement}, shows how these devices bear a remarkable risk of data breach. Later, the same authors contributed to similar experiments against VoIP phones~\cite{voip20}. Our work follows a similar methodology.

In 2022, a literature survey for understanding IoT security threats and challenges appeared~\cite{NATHN2022107997}. 
It is divided into three parts. The first part contains an analysis of the main threats and attacks, also analysing the IoT ecosystem from four perspectives: devices, internal network services, external network services and users. The second part contains a study on recent IoT malware attacks. Finally, the general security requirements and challenges to address the devised attack categories are discussed. This work was inspirational for our experiments.

PETIoT is a recent cyber Kill Chain (KC) specifically developed to guide VAPT sessions over IoT devices~\cite{petiot} with a focus on detecting their network vulnerabilities. The KC is demonstrated on the Tapo C200 IP camera by Tp-Link, enabling the discovery of three severe vulnerabilities: \textit{Improper neutralisation of inbound packets}; \textit{Insufficient entropy in encrypted notifications}; \textit{Cleartext transmission of video stream}. Each of the vulnerabilities found is exploited through different attack scenarios, and appropriate remediation measures are illustrated as possible fixes. We followed PETIoT strictly through our work, as can be seen by the steps discussed below.

In 2018, a penetration testing session on the \textit{Tradfri} smart bulb produced by \textit{IKEA} was reported~\cite{ikea_tradfri}. The article describes four different attacks. \textit{Hacking Smart Light Bulb via Bluetooth} allows the attacker to control the bulb. \textit{IKEA Tradfri Gateway Exploit} allows the attacker to compromise the ZigBee Tradfri gateway by managing to break the firmware update. \textit{Denial of Service Attack} allows the attacker to perform a DoS attack against the gateway by flooding it with UDP packets. \textit{Identity Spoofing} allows the attacker to impersonate the CoAP client and control the smart bulbs. This work is closely related to ours but revolves around the ZigBee protocol and does not achieve escalation to local Wi-Fi access.

\section{EXPERIMENT SETUP}\label{sec:1}
We begin defining the experiment setup, accordingly with the PETIoT kill chain. Our setup is non-invasive, as the smart bulb is not tampered with. It includes:
\begin{itemize}
	\item A Wi-Fi switch to provide local connectivity.
	\item A Smart bulb Tapo series L530 with Hardware Version 1.0.0 and Firmware Version 1.1.9.
	\item A Samsung smartphone running Android 11 and the Tapo app Version 2.8.14.
	\item An Ubuntu 22.04 machine with 5.15.0-47 kernel to run all software needed for the experiments.
\end{itemize}
To be able to use all the bulb features, it is necessary to create a Tapo account and login into it.
%
Depending on whether the smart bulb is already associated with a Tapo account or not, and the network configuration used, we identify three different setups. In all three setups, the smartphone has access to the Tapo account to which the smart bulb is associated, or to which the user wants to associate it.

\subsection{Setup A}\label{sec:scenario_A}
\begin{figure}[htbp]
\centerline{\includegraphics[scale=0.35]{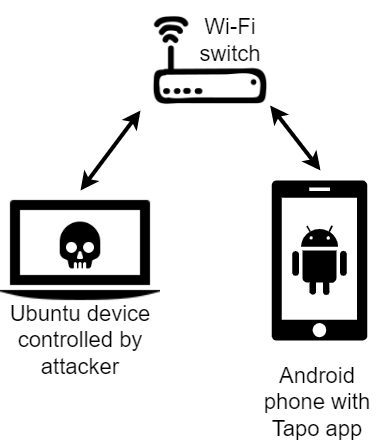}}
\caption{Setup A, network without a local smart bulb}
\label{fig:scenario_A}
\end{figure}

The context of the \textit{Setup A} is as follows. \begin{itemize}
    \item The user has previously associated her smart bulb with the Tapo app on her phone. During the association process, the user connects the smart bulb to a certain network $X$.
    \item The Ubuntu device and the user's phone are both connected to the same network $Y$ as shown in Fig.~\ref{fig:scenario_A} --- this is not necessarily the network $X$ to which the Tapo L530E is connected. 
\end{itemize}

This setup 
does not require the Ubuntu device to have access to the network to which the smart bulb is connected, but only to be connected to the same network as the Tapo app. Therefore, the user could, for example, connect to a public network to turn off a light that may have been mistakenly left on at home. 

\subsection{Setup B}\label{sec:scenario_B}
\begin{figure}[htbp]
    \centerline{\includegraphics[scale=0.35]{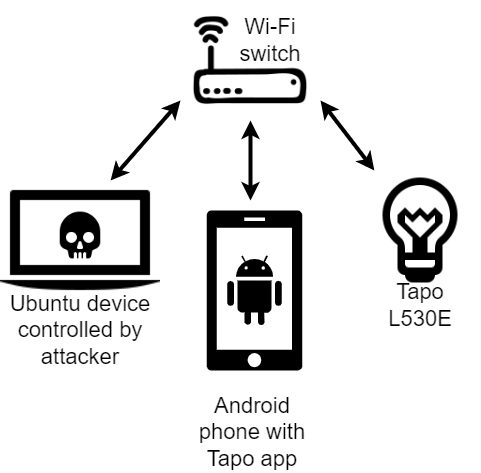}}
    \caption{Setup B, network with a configured smart bulb}
    \label{fig:scenario_B}
\end{figure}

The context of the \textit{Setup B} is as follows. \begin{itemize}
    \item The user has previously associated her smart bulb with the Tapo app on her phone. During the association process, the user connects the smart bulb to a certain network $X$.
    \item The Ubuntu device and the user's phone are both connected to the same network $X$ to which the smart bulb is connected, as shown in Fig.~\ref{fig:scenario_B}. 
    \item The Ubuntu device has complete control of the network $X$. It is able to carry out an \textit{ARP spoofing attack}, i.e., to intercept data frames, modify the network traffic or prevent it from reaching its intended recipient.
\end{itemize}

This setup requires the active presence of the smart bulb.
In addition, the setup requires the Ubuntu device to have access to the network to which the smart bulb is connected, typically the victim's local network. Therefore, all devices used are connected via Wi-Fi to the same access point.

\subsection{Setup C}\label{sec:scenario_C}
The context of the \textit{Setup C} is as follows. \begin{itemize}
    \item The user wants to associate the newly reset or not yet configured smart bulb with her Tapo account.
    \item The Tapo app (hence, the user) believes to be connected to the network $X$ created by the smart bulb, but it is actually connected to a network $Y$ controlled by the Ubuntu device.
    \item The Ubuntu device is connected both to the network created by the smart bulb and to the network controlled by itself.
\end{itemize}

\begin{figure}[htbp]
\centerline{\includegraphics[scale=0.35]{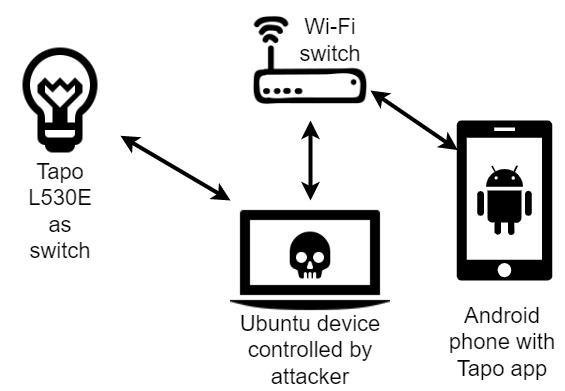}}
\caption{Setup C, network with a non-configured bulb}
\label{fig:scenario_C}
\end{figure}

This setup occurs when the smart bulb has been reset or has not been configured yet. As detailed in Section~\ref{sec:states_not_config}, the smart bulb starts a public access point to which the user must connect to complete the setup.
The network configuration is shown in Fig.~\ref{fig:scenario_C}.

The \textit{Wi-Fi deauthentication attack} allows the attacker to easily get \textit{Setup C}. A \textit{Wi-Fi deauthentication attack} is a type of denial-of-service attack that targets communications between a device and a Wi-Fi wireless access point~\cite{deauth_analysis}.
Thanks to this attack, the attacker can deauthenticate the smart bulb from the network to which it is connected, forcing the user to repeat its setup process.
The \textit{Wi-Fi deauthentication attack} requires the access point and connected devices not to use 802.11w or WPA3. Although in literature there are other theoretically feasible methods to perform the deauthentication attack anyway~\cite{deaut_Wi-Fi6_WPA3}, we did not test them. In \textit{Setup C}, we assume that the network to which the smart bulb is connected is a network where the deauthentication attack can be performed. One way to mount it would be to leverage \textit{Aircrack-ng}~\cite{aircrack_ng}, as described in a blog~\cite{Wi-Fi_deauth-tool}.
The attack requires the adversary to know the victim device's MAC address, however, it is possible to list all MAC addresses connected to nearby access points by simply listening to the traffic, so the adversary has a short and finite list of addresses to try.

After the victim resets the smart bulb, it enters setup mode.
In \textit{Setup C}, the Ubuntu device needs the Tapo app to connect to the network it controls and not to the network started by the smart bulb --- hence, the adversary performs another \textit{Wi-Fi deauthentication attack} to deauthenticate the phone running the Tapo app from the network started by the smart bulb, arguably inducing the victim to re-attempt to connect, eventually using the adversary-controlled network with the same SSID. As the network started by the smart bulb is an unprotected Wi-Fi 4 (802.11n) network, in this latter case the deauthentication attack will always work. Intuitively, this setup assumes that the Ubuntu device is close enough to the network started by the smart bulb to hear the emitted carrier.

%

\section{INFORMATION GATHERING} \label{sec:2}

After the setup we can proceed to the collection of information. The following tools are installed on the Ubuntu device:\begin{itemize}
	\item \textit{Ettercap}~\cite{ettercap}, a suite of tools to perform MITM attacks used on the Ubuntu device to control the network.
	\item \textit{Wireshark}~\cite{wireshark}, a network protocol analyser (or packet sniffer) used to capture and analyse traffic.
\end{itemize}
The smart bulb firmware source code is not publicly available. Therefore, the offensive activities are performed in black-box mode.

Thanks to the messages captured with \textit{Wireshark} it is possible to distinguish between three different types of communications. They differ according to the sender and the receiver of the various messages belonging to them. The three different types of communications we identified are: \textit{(a)} communications \textit{App-Cloud}, including all messages exchanged between the Tapo App and the Cloud Server; \textit{(b)} communications \textit{Bulb-App}, including all messages exchanged between the smart bulb and the Tapo App; \textit{(c)} communications \textit{Cloud-Bulb}, including all messages exchanged between the Cloud Server and the smart bulb.
%
When the phone running the Tapo app is not connected to the same network as the light bulb, communications happen through the cloud, combining \textit{App-Cloud} and \textit{Cloud-Bulb} communications -- otherwise, they take place locally with \textit{Bulb-App} communications.

All communications through the cloud, i.e., \textit{App-Cloud} and \textit{Cloud-Bulb}, are encrypted. They take place through the use of a secure TLS channel that ensures authenticity, integrity, and confidentiality of the messages, even though they are transmitted over the Internet. Instead, all \textit{Bulb-App} communications are exchanged via HTTP messages. Their payloads are encrypted using the AES128 block cipher in CBC mode. The initialisation vector and the cryptographic key used for this protocol are exchanged using the Tapo ``Symmetric Key Exchange Protocol'' (TSKEP), which is described below in Section \ref{sec:TSKEP}. This protocol only generates traffic over the local Wi-Fi network, not through the Internet.

Our offensive activities only target \textit{Bulb-App} communications, and ignore the others.

\section{TRAFFIC ANALYSIS}\label{sec:3}

Thanks to the information obtained during the previous phase, it is now possible to analyse the network traffic profitably.

The API exposed by the smart bulb is very similar to a \textit{Remote Procedure Call} (RPC). The JSON sent by the app to the smart bulb contains the name of the method to be invoked and its parameters. The responses sent by the smart bulb to the app contain an \textit{error code} representing the outcome of the operation (whether it was successfully executed or the error occurred) with an optional result. All JSONs are exchanged via the payloads of HTTP messages. 


Before the Tapo app starts using the API exposed by the bulb, it must perform two preliminary steps: locate the smart bulb within the network to which it is connected, and exchange a symmetric key with the smart bulb to encrypt messages. Therefore, the communication between smart bulb and Tapo app can be summarised in three macro-steps: 
\begin{itemize}
    \item \textit{Bulb Discovery} – it allows the Tapo app to locate the smart bulb within the local network and to get the smart bulb's current configuration.

    \item \textit{Tapo Symmetric Key Exchange Protocol} – executing the TSKEP protocol allows the Tapo app and smart bulb to exchange a symmetric key.

    \item \textit{Smart bulb usage} – it consists in using the actual smart bulb app protocol.
\end{itemize}
These macro-steps are described in detail below.

\subsection{Bulb Discovery}\label{sec:bulb-discovery}

\begin{figure}[htbp]
\centerline{\includegraphics[scale=0.4]{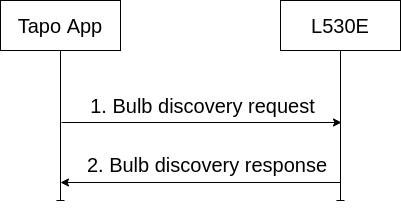}}
\caption{Tapo (local) Device dicovery}
\label{fig:bulb_discovery_step}
\end{figure}

Smart bulb configurations may change over time. For example, the IP address assigned to it by the DHCP server might have changed, it might start using a different encryption scheme after an update, or it might have just been reset. Before the Tapo app can start communicating with the smart bulb, it must locate the smart bulb within the local network and get its current configuration. To do this, the Tapo app connects to the UDP service listening on the $20002$ port of the smart bulb, as represented in Fig.~\ref{fig:bulb_discovery_step}. 

The payload format of such UDP messages is represented in Table~\ref{tab:udp_message_structure}. It can be explained as follows:

\begin{itemize}
    \item Bytes in positions [0:3] and [6:7] are static. Their values never change.
    
    \item The \textit{data length} field contains the length of the \textit{data} field, which contains the data exchanged between Tapo app and smart bulb. 
    When the data field is empty (i.e., when it has length 0) the bits of this field are all set to $0$.
    
    \item The \textit{nonce} field contains a 4-byte string randomly generated by the Tapo app. The string is inserted in the \textit{bulb discovery request} message from the Tapo app and quoted by the smart bulb in the \textit{bulb discovery response} message from the smart bulb. This allows the Tapo app to obtain guarantees about the freshness of the response received from the smart bulb. 
    
    \item Logically, the \textit{checksum} field must be excluded from the computation of the checksum itself — instead, a secret key of the same length is used in its place for the checksum computation. This secret key is hard-coded in both the Tapo app and the smart bulb, hence, the checksum acts like a Message Authentication Code (MAC).
    This field allows the receiver to understand that the message arrives intact as it is sent.
    The algorithm used for the \textit{checksum} calculation is the Cyclic Redundancy Check 32 (CRC-32), a checksum algorithm that hashes byte sequences to 32 bit values.
    
    \item The \textit{data} field contains different information depending on whether it is a \textit{bulb discovery request} or a \textit{bulb discovery response} and whether the smart bulb is configured or not.    
\end{itemize}

\begin{table}[htbp]
\begin{center}
    \begin{tabular}{cc|clll|}
        \cline{3-6} & & \multicolumn{4}{c|}{Octet} \\ \cline{3-6} & & \multicolumn{1}{c|}{0} & \multicolumn{1}{l|}{1} & \multicolumn{1}{l|}{2} & 3 \\ \hline \multicolumn{1}{|c|}{\begin{tabular}[c]{@{}c@{}}Offset\end{tabular}} & 0 & \multicolumn{1}{c|}{0x02} & \multicolumn{1}{l|}{0x00} & \multicolumn{1}{l|}{0x00} & 0x01 \\ \cline{2-6} \multicolumn{1}{|c|}{} & 4 & \multicolumn{2}{c|}{Data length} & \multicolumn{1}{l|}{0x11} & 0x00 \\ \cline{2-6} \multicolumn{1}{|c|}{} & 8 & \multicolumn{4}{c|}{Nonce} \\ \cline{2-6} \multicolumn{1}{|c|}{} & 12 & \multicolumn{4}{c|}{Checksum} \\ \cline{2-6} \multicolumn{1}{|c|}{} & ... & \multicolumn{4}{c|}{Data} \\ \hline 
    \end{tabular}
    \caption{Format of the payload of UDP messages}
    \label{tab:udp_message_structure}
    \end{center}
\end{table}

\longversion{
\subsubsection{Bulb Discovery Request}
This message is generated by the Tapo app for the smart bulb. At the network layer, the \textit{bulb discovery request} uses the IPv4 protocol. The source address of this request is the Tapo app IPv4 address, while the destination is the broadcast IPv4 address. At the transport layer, the \textit{bulb discovery request} uses the UDP protocol and the destination port is set to $20002$.

The message is broadcast, which means it is received by all devices connected to the network. Therefore, the goal of the message is not to locate a specific device within the network but to locate all devices compatible with the Tapo protocol connected to the network.
        
The \textit{data} field of the message can be empty or not. This depends on whether it is sent by the Tapo app to locate all devices already associated or all the ones that still have to be associated. In the former case, it will have a non-zero value, while in the latter it will be empty.

\subsubsection{Bulb Discovery Response}
This message is generated by the smart bulb for the Tapo app. At the network layer, the source address is the smart bulb address, while the destination address is the Tapo app address. At the transport layer, this message uses the UDP protocol too.

The \textit{nonce} field of the payload of this message contains the nonce of the \textit{bulb discovery request} message that it refers to. This will allow the Tapo app to get guarantees about the freshness of the message, and that it is not a replay of old configurations. The \textit{data} field of this message is never null and always reports a JSON containing the current smart bulb configurations.

The JSON contains the fields shown in Listing \ref{lst:discovery_response_json}, which we hereby explain:
\begin{lstlisting}[language=json,caption={UDP discovery response JSON},captionpos=b,label={lst:discovery_response_json}]
{
 "result": {
   "device_id": "bd1e...9348",
   "owner": "808e...c8e1",
   "device_type": 
     "SMART.TAPOBULB",
   "device_model": 
     "L530E Series",
   "ip": "192.168.1.151",
   "mac": "D2-AB-7F-1D-23-91",
   "factory_default": false,
   "is_support_iot_cloud": true,
   "mgt_encrypt_schm": {
     "is_support_https": false,
     "encrypt_type": "AES",
     "http_port": 80
   }
 },
 "error_code": 0 
}
\end{lstlisting}
\begin{itemize}
    \item The \textit{device\_id} and \textit{owner} fields are two 16-bytes strings transmitted in hexadecimal. These two strings represent the unique device id and the account id to which the device is associated, respectively. If the device is not yet associated with any account the \textit{owner} field contains a value that allows the Tapo app to understand that the smart bulb is not yet associated with any account.
    
    \item The \textit{device\_type} and \textit{device\_model} fields contain information for the Tapo app about the type of device that is sending the message. 

    \item The \textit{ip} and \textit{mac} fields contain the current address at both the network and device connection levels. The IP could change over time.
    
    \item The \textit{factory\_default} field allows the Tapo app to determine whether the device is already configured or yet to be configured. If the field is set to \texttt{true} then the smart bulb is not yet associated with any account. If it is set to \texttt{false} then the smart bulb is already associated with an account.
    
    \item The \textit{is\_support\_iot\_cloud} field allows the Tapo app to determine whether the device can be controlled via the cloud or not.

    \item The \textit{mgt\_encrypt\_schm} field provides the Tapo app with information about the encryption scheme supported by the smart bulb and the port where it is listening.
\end{itemize}
In our case, the smart bulb uses AES128, in the CBC mode, as encryption scheme. The Tapo app determines the type of communication to start depending on whether it is trying to associate or use the smart bulb.
}

\subsection{Tapo Symmetric Key Exchange Protocol}\label{sec:TSKEP}
\begin{figure}[htbp]
\centerline{\includegraphics[scale=0.35]{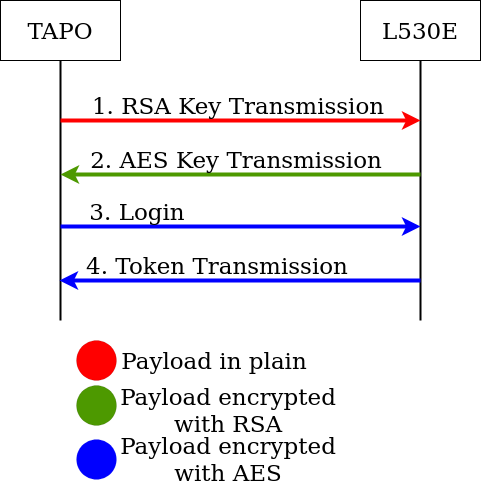}}
\caption{Tapo Symmetric Key Exchange Protocol}
\label{fig:tskep_flow}
\end{figure}

This step uses the RPC API exposed by the smart bulb on port $80$. All exchanged messages are HTTP messages, so at the transport layer, only the TCP protocol is used. The TSKEP, represented in Fig.~\ref{fig:tskep_flow}, 
allows Tapo app and smart bulb to exchange a 128-bit AES key and an IV to encrypt the payload of various HTTP messages. The protocol consists of four different messages:\begin{enumerate}
    \item RSA public key transmission
    \item AES key transmission
    \item Login
    \item Token transmission
\end{enumerate}

At the end of this phase, Tapo app and smart bulb get a short-term shared secret to encrypt all subsequent traffic. 
Thanks to the credential stored during the setup phase, for the smart bulb the secret is also authenticated.

\longversion{
\subsubsection{RSA public key transmission}
The Tapo app starts the symmetric key exchange protocol by making an HTTP request using the POST method. From the JSON contained in the payload, shown in Listing \ref{lst:rsa_key_transmission}, we can see that this first request invokes the \textit{handshake} method on the smart bulb and passes as a parameter a public RSA  key of which the Tapo app knows the private one.
\begin{lstlisting}[language=http,caption={RSA public key transmission message JSON},captionpos=b,label={lst:rsa_key_transmission}]
  POST /app HTTP/1.1
  Referer: http://192.168.1.151:80
  Accept: application/json
  requestByApp: true
  Content-Type: application/json; charset=UTF-8
  Content-Length: 338
  Host: 192.168.1.151
  Connection: Keep-Alive
  Accept-Encoding: gzip
  User-Agent: okhttp/3.12.13
  
  {
   "method":"handshake",
   "params":{ 
    "key":"-----BEGIN PUBLIC KEY-----\nMIGfMA0GCSqGSIb3D ... CwIDAQAB\n-----END PUBLIC KEY-----\n"
   },
   "requestTimeMils":0
  }
\end{lstlisting}

\medbreak

\subsubsection{AES key transmission}
When the Tapo app invokes the \textit{handshake} method, the smart bulb generates a 16-byte symmetric key, a 16-byte IV, and a session cookie. The symmetric key and the IV are used by Tapo app and smart bulb to encrypt the payloads of all subsequent messages using AES128 in the CBC operation mode. The cookie generated by the smart bulb is valid for 24 hours, and it is included by the Tapo app in every subsequent message. Hence, this allows the smart bulb to associate the message with the correct key and the correct IV to decrypt.
An example of this protocol message is shown in Listing  \ref{lst:aes_iv_key_transmission}.
\begin{lstlisting}[language=http,caption={AES key transmission JSON},captionpos=b,label={lst:aes_iv_key_transmission}]
  HTTP/1.1 200 OK
  Content-Type: application/json;charset=UTF-8
  Content-Length: 208
  Connection: close
  Set-Cookie: TP_SESSIONID=157A987844588259EC490749B8494549;TIMEOUT=1440
  
  {
   "error_code":0,
   "result":{
    "key":"l1adeEm ... S5aZZxX1hA="
    }
  }
\end{lstlisting}
The AES128 key and IV are concatenated and encrypted using the Tapo app’s RSA public key. Moreover, the base64 encoding of the resulting ciphertext is included in the \textit{key} field of the JSON returned to the Tapo app by the smart bulb. The cookie is included in the \textit{Set-Cookie} HTTP response header.

\subsubsection{Login} 
Only accounts with which the smart bulb is shared must be able to send commands to it. For this reason, the Tapo app needs to authenticate to the smart bulb. This is done with the \textit{Login} message. In this message, the Tapo app sends the base64 encoding of the username and the password of the user's account to the smart bulb. The username is the SHA1 hash of the victim user’s email.

The \textit{login} message invokes the \textit{securePassthrough} method on the smart bulb. When this method is invoked, the smart bulb expects the passed parameters to be encrypted using the AES128 key and IV previously exchanged.
An example of this protocol message is shown in Listing~\ref{lst:login_msg}.
\begin{lstlisting}[language=http,caption={Login JSON},captionpos=b,label={lst:login_msg}]
  POST /app HTTP/1.1
  Referer: http://192.168.1.151:80
  Accept: application/json
  requestByApp: true
  Content-Type: application/json; charset=UTF-8
  Content-Length: 276
  Host: 192.168.1.151
  Connection: Keep-Alive
  Accept-Encoding: gzip
  Cookie: TP_SESSIONID=157A987844588259EC490749B8494549
  User-Agent: okhttp/3.12.13
  
  {
   "method":"securePassthrough",
   "params":{
    "request":"vF+KC5IW5Kh ... ZNxa86G7v9p szvFsYSFWNRRLUHm5tPAA==\n"
    }
  }
\end{lstlisting}
The parameter passed in the \textit{request} field is the base64 encoding of the RSA encryption of a JSON containing the invocation of another method. In the case of the \textit{login} message, the method invoked is \textit{login\_device}. The decryption of the \textit{request} field produces the JSON shown in Listing~\ref{lst:plaintext_of_login_msg}: 
\begin{lstlisting}[language=json,caption={Content of the \textit{params} field in plaintext},captionpos=b,label={lst:plaintext_of_login_msg}]
  {
   "method":"login_device",
   "params":{
    "password":" . . . . ",
    "username":" . . . . "
   },
   "requestTimeMils":0
  }
\end{lstlisting}

\subsubsection{Token transmission} 
The smart bulb authenticates the Tapo app and generates and returns a token. The token is used by the Tapo app to authenticate to the smart bulb without having to send a username and password every time. 
The token generated by the smart bulb is encrypted with the symmetric key, previously exchanged, and inserted inside the HTTP message shown in Listing~\ref{lst:token_transmission}.
\begin{lstlisting}[language=http,caption={Token transmission message},captionpos=b,label={lst:token_transmission}]
  HTTP/1.1 200 OK
  Content-Type: application/json;charset=UTF-8
  Content-Length: 149
  
  {
    "error_code":0,"result":{
      "response":"1i7xWcMMNbf7MQ95ciVEY57rs4T+3 ... fuTwa6J7OnW1aRwAxlRKpkM="
    }
  }
\end{lstlisting}
When the method invoked by the Tapo app is \textit{securePassthrough}, the \textit{result} field of the smart bulb's response JSON contains the base64 encoding of a cryptotext. The key and IV used by the smart bulb to get the encryption are the same as the Tapo app used in the request. 
}
\subsection{Smart Bulb Setup}
This section focuses on the two possible configurations for the smart bulb and specifies them in depth following our traffic analysis efforts.
 
\subsubsection{Smart bulb turned on and not configured}\label{sec:states_not_config}
Before a smart bulb Tapo L530E can be used, it must be associated with a Tapo account. There are two reasons why a smart bulb may not be associated with any accounts: because it has been reset, or it has not been configured yet. In this Section, we discuss the process of associating the smart bulb with a Tapo account. 

An unconfigured or newly reset smart bulb starts a public access point with SSID \textit{Tapo\_Bulb\_XXXX}, where $XXXX$ are four decimal places. The smart bulb also acts as a switch within the network it generates. In order to configure it and associate it with their Tapo account, the user must connect to the Wi-Fi network started by the smart bulb itself.

After that, the Tapo app tries to locate the smart bulb. To do this, it starts sending \textit{bulb discovery request} messages to broadcast. In this case, the \textit{data} field of these messages is empty.
After the identification of the smart bulb, the Tapo app starts the TSKEP protocol with it. 
The values set in the login message as \texttt{username} and \texttt{password} are fixed values that the Tapo app uses every time it configures a new device.

Once the symmetric key is obtained, Tapo app sends to the smart bulb the SSID and the password of the Wi-Fi network to which the smart bulb must connect. The Tapo app also sends the credentials of the Tapo account to which it must be associated. 
The credentials are then stored by the smart bulb. Through these credentials, smart bulb is able to authenticate all subsequent requests of Tapo app.
At this point, the smart bulb turns its access point off and connects to the specified Wi-Fi network. The smart bulb starts communicating with the cloud server to complete its setup. Hence, the Wi-Fi network to which the smart bulb connects must have Internet access. 


\subsubsection{Smart bulb already configured}
Let us now consider the case where the smart bulb is associated with a Tapo account and it is ready to be used. As mentioned before, it can be controlled either locally or remotely via the Tapo cloud. To do so, the Tapo application  
%
initially tries to locate the smart bulb within the network with a \textit{bulb discovery request} message. If it detects it (i.e., it gets a response) then the interaction happens locally via Bulb-App communications, which are the subject of our analysis.
Otherwise, if the Tapo app does not receive any valid \textit{bulb discovery response}, then it tries to check the smart bulb remotely. If the smart bulb is not detectable even remotely then it is determined offline.

\section{VULNERABILITY ASSESSMENT}\label{sec:4}
The assessment following the information gathered so far highlights four vulnerabilities.

\paragraph{Vulnerability 1 – Lack of the smart bulb authentication with the Tapo app.}
\textit{Improper Authentication}~\cite{CWE-287} in Tapo L503E allows an adjacent attacker to impersonate the Tapo L530E with the Tapo app during the TSKEP step.

In the TSKEP step, unlike the \textit{Bulb Discovery} step, the protocol used to exchange the session key does not give the Tapo app any evidence of its peer's identity. Hence, an attacker is able to authenticate to the Tapo app as the Tapo L530E or as another device: in fact, this vulnerability is present in all Tapo smart devices that use the TSKEP protocol.

\longversion{Figure ~\ref{vuln 1 cvss} shows the \textit{CVSS v3.1} score that we calculate, i.e. 8.8, meaning \textit{High} severity.} 
The \textit{CVSS v3.1} score that we calculate is 8.8, meaning \textit{High} severity. Precisely: Attack Vector: Adjacent; Attack Complexity: Low; Privileges Required: None; User Interaction: Required; Scope: Changed; Confidentiality: High; Integrity: High; Availability: High.
In particular, Attack Complexity is low because the attacker could impersonate the bulb by implementing the protocol messages to respond to the calling app. Following that, he could obtain the user password on the Tapo app, then fully impersonate the user and manipulate at will any target Tapo device of the same user. Precisely, by impersonating the bulb at setup time as explained above, the attacker would receive the victim's Wi-Fi SSID and password from the Tapo app, so that he could then 
impersonate the user by her password at each session with the target device, which could be any Tapo device of the user's. The attacker could also obtain the device-chosen session key, which he could then relay to the user's genuine app and effectively interpose.

\longversion{\begin{figure}[htbp]
\centerline{\includegraphics[scale=0.4]{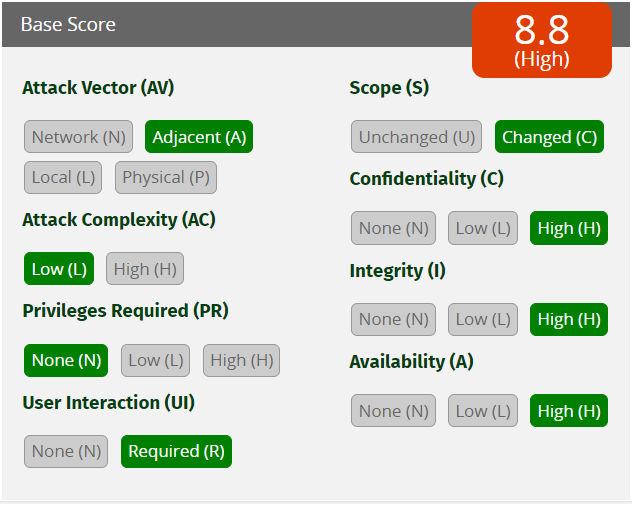}}
\caption{CVSS calculated for Vulnerability 1 – Lack of the smart bulb authentication with the Tapo app}
\label{vuln 1 cvss}
\end{figure}}

\paragraph{Vulnerability 2 – Hard-coded short checksum shared secret.}
\textit{Protection Mechanism Failure}~\cite{CWE-693} in Tapo L503E allows an adjacent attacker to obtain the secret used for authentication during the \textit{Bulb Discovery} phase.

The shared secret used for \textit{Bulb Discovery}'s messages authentication is short and hard-coded both in the Tapo app and in the Tapo L530E. Therefore, it can be obtained in the following ways:
\begin{enumerate}
    \item Brueforcing, because of its shortness.
    \item Decompiling the Tapo app.
\end{enumerate}

\longversion{Figure ~\ref{vuln 2 cvss} shows the \textit{CVSS v3.1} score that we calculate, i.e. 7.6, meaning \textit{High} severity.}
The \textit{CVSS v3.1} score that we calculate is 7.6, meaning \textit{High} severity. Precisely: Attack Vector: Adjacent; Attack Complexity: Low; Privileges Required: None; User Interaction: Required; Scope: Unchanged; Confidentiality: Low; Integrity: High; Availability: High.
The knowledge of the shared secret provides to the attacker the ability to edit and to create \textit{Bulb Discovery} messages. Specifically, the attacker is able to generate fake \textit{bulb discovery requests} to locate all smart bulbs, or generally Tapo devices using the same protocol, connected within the same network. Meanwhile, the ability to edit valid messages allows the attacker to edit the \textit{bulb discovery response} messages sent from the smart bulb to the Tapo app.

\longversion{\begin{figure}[htbp]
\centerline{\includegraphics[scale=0.4]{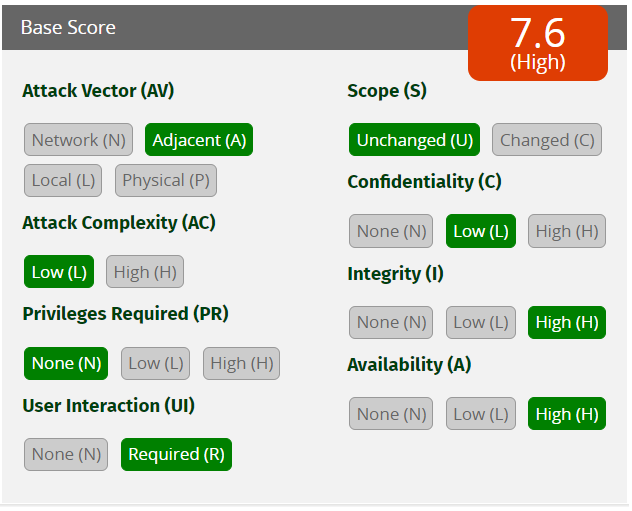}}
\caption{CVSS calculated for Vulnerability 2 – Hard-coded short checksum shared secret}
\label{vuln 2 cvss}
\end{figure}}

 \paragraph{Vulnerability 3 – Lack of randomness during symmetric encryption.}
 \textit{Use of a Cryptographic Primitive with a Risky Implementation}~\cite{CWE-1240} in Tapo L503E allows an adjacent attacker to make the cryptographic scheme deterministic.

The IV used in AES128-CBC scheme is generated together with the key and remains unchanged for the entire life of the key. Therefore, the encryption of the same messages produces the same ciphertext.

\longversion{Figure ~\ref{vuln 3 cvss} shows the \textit{CVSS v3.1} score that we calculate, i.e. 4.6, meaning \textit{Medium} severity.}
The \textit{CVSS v3.1} score that we calculate is 4.6, meaning \textit{Medium} severity. Precisely: Attack Vector: Adjacent; Attack Complexity: Low; Privileges Required: None; User Interaction: Required; Scope: Unchanged; Confidentiality: Low; Integrity: None; Availability: Low.
When the user interacts with the device thereby generating traffic, the attacker can distinguish repeated messages without deciphering them, yet infer what messages lead to what consequences, such as turning the bulb off. He could then manipulate the bulb by repeating those messages.

\longversion{\begin{figure}[htbp]
\centerline{\includegraphics[scale=0.4]{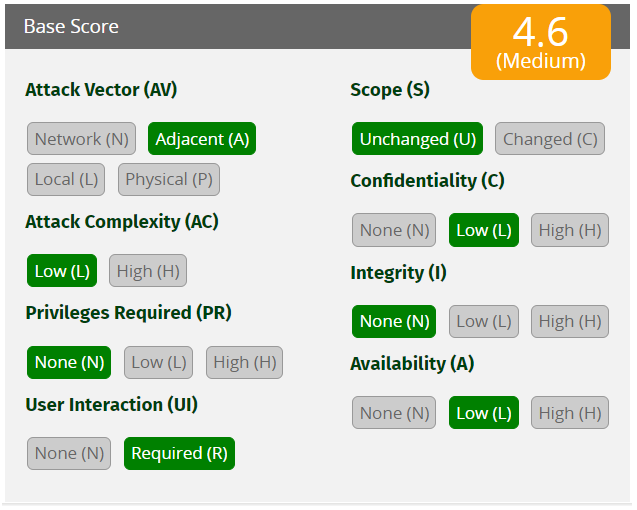}}
\caption{CVSS calculated for Vulnerability 3 – Lack of randomness during symmetric encryption}
\label{vuln 3 cvss}
\end{figure}}

 \paragraph{Vulnerability 4 – Insufficient message freshness.}
 \textit{Predictable from Observable State}~\cite{CWE-341} in Tapo L503E allows an adjacent attacker to replay messages that are considered valid both from the Tapo L530E and the Tapo app.

 The smart bulb and the Tapo app do not check the freshness or the duplicity of the received messages. They only check that the session key with which the messages are encrypted is still valid, i.e., not older than 24 hours.
 
 \longversion{Figure ~\ref{vuln 4 cvss} shows the \textit{CVSS v3.1} score that we calculate, i.e. 5.7, meaning \textit{Medium} severity.}
The \textit{CVSS v3.1} score that we calculate is 5.7, meaning \textit{Medium} severity. Precisely: Attack Vector: Adjacent; Attack Complexity: Low; Privileges Required: None; User Interaction: Required; Scope: Unchanged; Confidentiality: None; Integrity: None; Availability: High.
Similarly to the previous vulnerability, the attacker can leverage user-generated traffic, this time to replay messages that both the bulb and the app will certainly accept because of the lack of appropriate freshness measures.

\longversion{\begin{figure}[htbp]
\centerline{\includegraphics[scale=0.4]{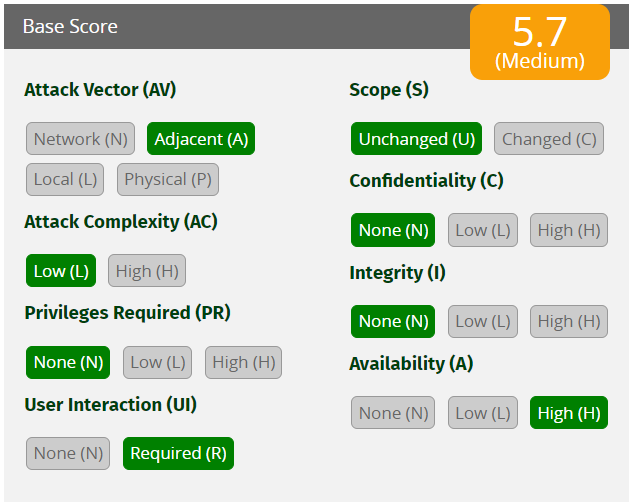}}
\caption{CVSS calculated for Vulnerability 4 – Insufficient message freshness}
\label{vuln 4 cvss}
\end{figure}}

\section{EXPLOITATION}\label{sec:5}
This Section shows how an attacker can exploit the vulnerabilities we found in a real environment. We show $5$ attack scenarios, in which the attacker exploits one or more vulnerabilities to achieve their malicious goals. We validated each attack scenario by manually executing the steps illustrated in the forthcoming sections, hence all reported attacks are feasible in practice. As noted above through the CVSS scores, their likelihood is determined by the Adjacent Attack Vector, the Low Attack Complexity, the No Privileges Required and the Required User Interaction.

\subsection{Attack scenario 1 - Fake Bulb Discovery messages generation}
In this experiment we exploited:\begin{itemize}
    \item \textit{Vulnerability 2}, with the goal of getting the ability to create fake \textit{Bulb Discovery} messages.
\end{itemize}

This experiment can be conducted in every scenario presented in Section~\ref{sec:1}. For this attack, it is first necessary to get hold of a UDP message. 
In our case, we have chosen to use a real \textit{bulb discovery request} message, because they are easy to get. These messages are broadcast by the Tapo app every time it is opened, regardless of the network to which it is connected. To get a valid one, we just capture the traffic using Wireshark and use a filter to extract all UDP messages sent in broadcast 
e.g., by using the filter \texttt{udp \&\& ip.dst == 255.255.255.255}.
Once a UDP message is obtained, we can perform an offline brute-force attack to get the secret shared between the Tapo app and the smart bulb. At this point, neither the smart bulb nor the Tapo app needs to be active anymore to complete the attack. 
In our setup, the brute-force attack always succeeded in 140 minutes on average.

This grants the adversary the ability to create fake \textit{bulb discovery request} and \textit{response} messages: the former allows the attacker to identify all Tapo devices that use the same protocol and the same key, on any network he connects to, while the latter allows the attacker to respond to the Tapo app's request messages with false configurations. 

\subsection{Attack scenario 2 - Password exfiltration from Tapo user account}
In this experiment we exploited, in order:\begin{itemize}
    \item \textit{Vulnerability 2}, with the goal of getting the ability to create fake \textit{bulb discovery response} messages,
    \item \textit{Vulnerability 1}, with the goal of getting the ability to authenticate as the Tapo L530E to the Tapo app.
\end{itemize}

\begin{figure}[htbp]
\centerline{\includegraphics[scale=0.35]{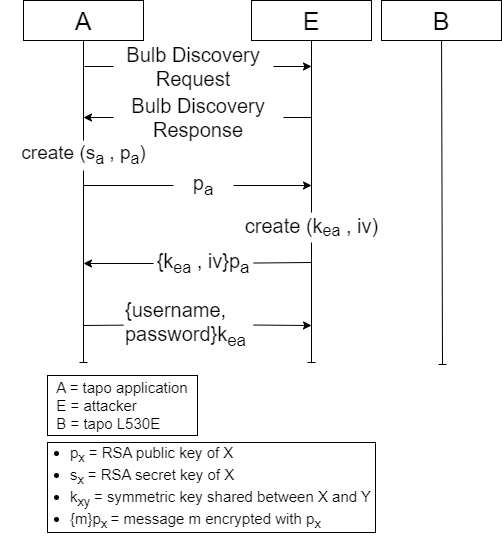}}
\caption{Sequence chart for the attacker's impersonation of the bulb}
\label{fig:steal_password}
\end{figure}

The context in which we conduct the experiment is Setup A described in Section~\ref{sec:scenario_A}. 
To carry out this attack, the adversary needs to obtain the \textit{ownerId} of the victim Tapo account, and the UDP port to which the victim sends her \textit{bulb discovery request} messages (in our case, port $20002$).
%
The attack diagram is shown in \textit{Fig.~\ref{fig:steal_password}}.

The exploitation begins the moment the victim opens her own Tapo app. When the app is open, it starts broadcasting \textit{bulb discovery request} messages. 
During the attack, the attacker exploits his ability to create fake \textit{bulb discovery response} messages to respond to various \textit{bulb discovery request} from the victim. The attacker sets the \textit{bulb discovery response} messages fields as shown in Listing \ref{lst:attacco 2}:\begin{itemize}
    \item It sets the \textit{owner} field to the \textit{ownerId} of the victim. This is to make the Tapo app think there is a device of its own on the network to start TSKEP.
    \item It sets the \textit{ip} and \textit{port} fields to point to an adversary-controlled server.
\end{itemize}

\begin{lstlisting}[language=json,caption={JSON attack scenario 2},captionpos=b,label={lst:attacco 2}]
{
 "result": {
   "device_id": "bd1e...9348",
   "owner": "Victim's ownerId",
   "device_type": 
     "SMART.TAPOBULB",
   "device_model": 
     "L530E Series",
   "ip": "Attacker's IP",
   "mac": "Attacker's MAC",
   "factory_default": false,
   "is_support_iot_cloud": true,
   "mgt_encrypt_schm": {
     "is_support_https": false,
     "encrypt_type": "AES",
     "http_port": 80
   }
 },
 "error_code": 0 
}
\end{lstlisting}

After receiving the response, the Tapo app thinks that it has successfully completed the \textit{Bulb Discovery} phase by locating its own device within the network. Therefore, it starts the TSKEP protocol with the attacking device. 
Because of vulnerability 1, the TSKEP protocol does not give the Tapo app any evidence about the identity of the interlocutor. For this reason, the Tapo app assumes that the newly received key is shared with an associated device, while it is shared with the attacker instead. Hence, the adversary is able to decrypt the \textit{Login} message of the TSKEP protocol and get the password and the hash of the email of the victim’s Tapo account.

The attack can be summarised as follows:\begin{itemize}
        \item The attacker gets the \textit{Bulb Discovery} shared key and creates fake \textit{bulb discovery response} messages. Therefore, the authentication of the \textit{bulb discovery response} message fails. 
        
        \item The Tapo app executes the TSKEP protocol with the attacker instead of the smart bulb. Therefore, authentication of the smart bulb with the Tapo app fails. This results in an \textit{integrity loss}.
        
        \item The Tapo app shares the key with the attacker, hence the distribution of the session key fails. This results in an \textit{availability loss}.

        \item The attacker can violate the confidentiality of the messages and get the password and the hash of the email of the victim’s Tapo account. This results in a \textit{confidentiality loss}.
\end{itemize}

\subsection{Attack scenario 3 - MITM attack with a configured Tapo L530E}\label{sec:attack_scenario_3}
In this experiment we exploited:\begin{itemize}
    \item \textit{Vulnerability 1}, with the goal of getting the ability to authenticate as the Tapo L530E to the Tapo app.
\end{itemize}



\begin{figure}[htbp]
\centerline{\includegraphics[scale=0.35]{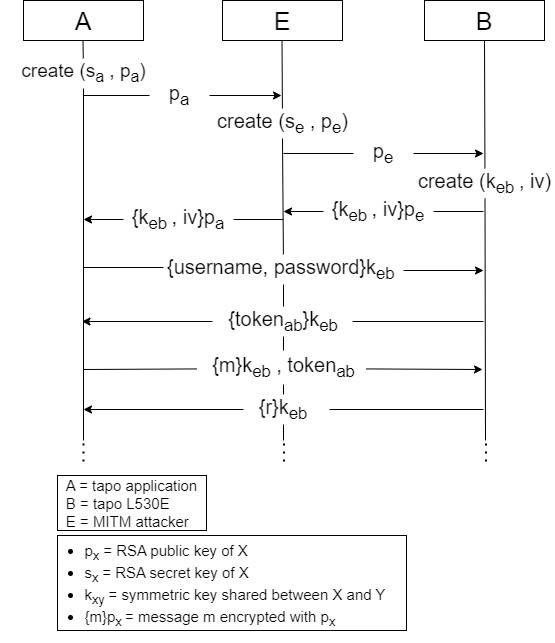}}
\caption{Sequence chart for the attacker's MITM between app and bulb}
\label{fig:break_confidentiality}
\end{figure}

The context in which we conduct the experiment is Setup B described in Section~\ref{sec:scenario_B}.
The attacker makes independent connections with the victims and relays messages between them to make them believe they are talking directly to each other over a private connection. 
When the Tapo app starts the TSKEP with the smart bulb, the attacker intercepts the \textit{RSA key transmission} message and blocks its reception from the smart bulb. 
In parallel, he starts a new session with the smart bulb from which he gets a new session key. The session key received by the attacker from the smart bulb is then encrypted with the previously received RSA public key and sent to the Tapo app. 
Due to vulnerability 1, the Tapo app expects to share the received key only with the smart bulb, but it is sharing it with the adversary instead. Hence, the attacker now has the key that Tapo app and smart bulb use to encrypt all subsequent communication messages. Therefore, he is capable of deciphering them and violating their confidentiality and integrity, for example by decrypting the messages, modifying their contents, re-encrypting and then forwarding them. This is summarised in the attack diagram shown in \textit{Fig.~\ref{fig:break_confidentiality}}.

Hence, the attack consists of the following steps:

\noindent\begin{itemize}
        \item The Tapo app executes the protocol with the attacker instead of the smart bulb, therefore, the authentication process fails. 
 
        \item The Tapo app shares the key with the attacker and not with the smart bulb, hence, the distribution of the session key fails.
 
        \item The attacker authenticates the key shared with the smart bulb thanks to the credentials received from the Tapo app, therefore, the authentication of the Tapo app with the smart bulb fails.
	
        \item The attacker relays messages between the two sessions to make Tapo app and smart bulb believe they are talking to each other, hence, the \textit{confidentiality} and the \textit{integrity} of messages is lost.
\end{itemize} 

\subsection{Attack scenario 4 - Replay attack with the Smart bulb as victim}
In this experiment we exploited:\begin{itemize}
    \item \textit{Vulnerability 4}, with the goal of getting the ability to replay both old and replicated messages.
\end{itemize}
%
The context in which we conduct the experiment is Setup B described in Section~\ref{sec:scenario_B}.
This attack scenario is divided into three phases. Specifically, in this case, it is also necessary that the attacker has a line of sight on the bulb to complete the attack.

During the first phase, \textit{Wireshark} is used to \textit{sniff} the traffic of a \textit{Bulb-App} communication. 
During this communication, the app sends to the smart bulb both \textit{get} messages, to request the value of some status parameters, and \textit{set} messages, to request the smart bulb to change the value of some of its internal parameters. 
During the experiment, we are not aware of the symmetric key used by both the smart bulb and the Tapo app to encrypt the messages. 

During the second phase of the experiment, we replicate all messages we previously captured. Hence, for each message we determine whether it was a \textit{get} or \textit{set} message simply by observing how the bulb behaves after each message. For every \textit{set} message, we take note of the change that it caused on the light bulb.

During the third phase, we arbitrarily replicate the \textit{set} messages to the smart bulb, managing to make changes to its internal state, without having a Tapo account associated with it. Messages continue to be accepted by the smart bulb until the session key with which they are encrypted expires.

\subsection{Attack scenario 5 - MITM attack with an unconfigured Tapo L530E}
In this experiment we exploited:

\begin{itemize}
    \item \textit{Vulnerability 1}, with the goal of getting the ability to authenticate as the Tapo L530E to the Tapo app.
\end{itemize}
%

The context in which we conduct the experiment is Setup C described in Section~\ref{sec:scenario_C}.
This attack scenario exploits the fact that not only the Tapo app, but anyone (including the attacker), can connect to the Wi-Fi network started by the smart bulb during the setup phase.
It is important that the attacking device acts as a bridge between the two networks. The attacker must flow all \textit{bulb discovery request} messages from the network it controls to the smart bulb network, and vice versa for \textit{bulb discovery response} messages. Otherwise, the Tapo app would not be able to detect the smart bulb and therefore would never try to start the TSKEP with it.
%
Subsequently, \textit{vulnerability 1} is exploited in the same way shown in Section \textit{Attack Scenario \ref{sec:attack_scenario_3}}, so the adversary is able to violate the confidentiality of the session key between the smart bulb and the Tapo app.

At some point in the communication, the Tapo app sends the JSON shown in Listing~\ref{lst:JSON_attack_scenario_5}:
\begin{lstlisting}[language=json,caption={JSON attack scenario 5},captionpos=b,label={lst:JSON_attack_scenario_5}]
{
 "method":"set_qs_info",
 "params":{
  "account"{
   "password":"Tapo password",
   "username":"Tapo email"
  },
  "extra_info":{"specs":"EU"},
  "time":{"region":"Europe/Rome","time_diff":60,"timestamp":1660032435},
  "wireless":{
   "key_type":"wpa2_psk",
   "password":"Wi-Fi password",
   "ssid":"ssid Wi-Fi"}
  },
  "requestTimeMils":1660032438365,
  "terminalUUID":"..."}'
\end{lstlisting}

Because usernames, Tapo passwords, SSIDs, and Wi-Fi passwords are sent in base64 encoding, the attacker is able to decode and steal them.

\section{FIXING}\label{sec:6}

This Section outlines possible mitigations for the identified vulnerabilities, marking the last step of the PETIoT kill chain.
Most measures consist of simple modifications to the relevant protocols, in particular to strengthen the cryptographic measures.
These can be implemented via software updates to be pushed to the affected devices via Internet, so we believe these fixes to be easily deployable with the already existing update procedures

\paragraph{Fix for Vulnerability 1.}
This vulnerability is the most complex and dangerous. It is not easy to find a simple fix to it because the protocol should be widely revised.
Our proposed fix requires the smart bulb to sign the message of \textit{AES key transmission} with an asymmetric, private key. 
The validity of that key as to belong to the bulb could be verified by the app via a digital certificate to retrieve from the Cloud Server during the association of the bulb with the app.
Of course, such a certificate should chain up to a root certificate to be securely stored with the app.
All this would allow the app to get evidence about the authenticity of the response, i.e., that the response really comes from the smart bulb.
In consequence, the app will eventually store all certificates of the associated devices.

\paragraph{Fix for Vulnerability 2.}
One possible solution to fix this vulnerability is the active presence of the cloud server. This entity should periodically assign each Tapo account a fresh key to use when calculating the checksum within \textit{Bulb Discovery} messages. The key assigned to a Tapo account should then be communicated to all devices associated with it. 
The benefits of the fix can be summarized as follows:\begin{itemize}
    \item The key is not hard-coded, so the attacker would no longer be able to get it by decompiling the Tapo app or the firmware of a Tapo device.

    \item Each account has its own key, therefore, compromising a Tapo account, or a key, would not result in compromising the keys of other Tapo accounts.

    \item The key should be long and random enough by current standards so that brute-force attacks would not be profitable anymore.
    
    \item The key is always fresh, so even if an attacker were to get the key of a Tapo account, the latter would not be compromised forever, but only until the validity of the stolen key expires and the cloud server assigns a new key to it.
\end{itemize}
It would also be useful to use a collision-resistant cryptographic hash function for the checksum. Examples of cryptographic hash functions are SHA-224 or SHA3-224.

 \paragraph{Fix for Vulnerability 3.}
 This vulnerability can be fixed by making the IV dynamic, i.e., using different IVs to encrypt different messages. This should be done by both the Tapo app and the Tapo L530E.
The IV used to encrypt the JSON contained in the \textit{params} field could then be included as a field in the plain part of JSON contained in \textit{Bulb-App} communications.

 \paragraph{Fix for Vulnerability 4.}
The timestamp containing the message creation moment included in JSON should be verified by smart bulb and the Tapo app. Checking the creation timestamp would prevent recent messages from being passed off as fresh by an attacker.
In addition, the various messages exchanged should contain a sequence number, which would prevent duplicate messages.

\section{CONCLUSIONS}\label{sec:concl}
We identified four vulnerabilities in the Tapo L530E, which we were able to practically exploit in five different attack scenarios with varying impacts on the users' security, privacy and safety. Of the four vulnerabilities, two are of High severity and two are of Medium severity, according to their CVSS score.

Overall, we observe that the experiment setup had to be designed with care due to the three scenarios that were possible. Following that, the information gathering step was rather large and complicated, much more than it could be reported in this paper due to space constraints. The vulnerability assessment was very surprising. For example, while deauthentication is routinely possible, we were not prepared to discover passwords in the clear and weak cryptography. Exploiting the vulnerabilities was moderately challenging but devising appropriate fixes was harder. 

One way to interpret such findings could be that ``small'' IoT devices may have raised insufficient cybersecurity attention thus far, i.e., insufficient cybersecurity measures due to a preconception that they may not be worth hacking or exploiting. Our work pins down this preconception as wrong, at least because the scope of our attacks expands onto all devices of the Tapo family a victim may use and, most importantly, potentially onto the entire victim's Wi-Fi network, which the attacker is enabled to penetrate.

While more and more experiments will certainly follow on similar bulbs and other inexpensive devices, we argue that the evidence we have gathered thus far is sufficient to call for a fuller application of a zero trust model to the IoT domain. With dozens of years of cybersecurity experience accumulated by the international community thus far, it should be possible to find affordable ways to achieve that in due course.

\bibliographystyle{apalike}
{\small
\bibliography{main}}

\begin{thebibliography}{}

\bibitem[Alopix, 2023]{Wi-Fi_deauth-tool}
Alopix (2023).
\newblock Wifi deauth attack in 2 minutes!
\newblock
  \url{https://systemweakness.com/wifi-deauth-attack-in-2-minutes-d1fb55112305}.

\bibitem[Bella and Biondi, 2019]{strive19}
Bella, G. and Biondi, P. (2019).
\newblock You overtrust your printer.
\newblock In {\em Proc. of the 2nd International Workshop on Safety, securiTy,
  and pRivacy In automotiVe systEms {(STRIVE'19)}}, LNCS 11699, pages 264--274.
  Springer.

\bibitem[Bella et~al., 2023]{petiot}
Bella, G., Biondi, P., Bognanni, S., and Esposito, S. (2023.).
\newblock {PETIoT: PEnetration Testing the Internet of Things}.
\newblock {\em Elsevier Internet of Things}, 22.

\bibitem[Biondi et~al., 2020]{voip20}
Biondi, P., Bognanni, S., and Bella, G. (2020).
\newblock {VoIP Can Still Be Exploited --- Badly}.
\newblock In {\em Proc. of the 5th International Conference on Fog and Mobile
  Edge Computing, {(FMEC'20)}}, pages 237--243. {IEEE}.

\bibitem[Dalvi et~al., 2018]{ikea_tradfri}
Dalvi, A., Maddala, S., and Suvarna, D. (2018).
\newblock Threat modelling of smart light bulb.
\newblock In {\em 2018 Fourth International Conference on Computing
  Communication Control and Automation (ICCUBEA)}, pages 1--4.

\bibitem[d'Otreppe~de Bouvette, 2023]{aircrack_ng}
d'Otreppe~de Bouvette, T. (2023).
\newblock {Aircrack-ng}.
\newblock \url{https://www.aircrack-ng.org/}.

\bibitem[Ettercap, 2023]{ettercap}
Ettercap (2023).
\newblock {Ettercap project}.
\newblock \url{https://www.ettercap-project.org/}.

\bibitem[Howarth, 2023]{statistics}
Howarth, J. (2023).
\newblock 80+ amazing iot statistics (2023-2030).
\newblock \url{https://explodingtopics.com/blog/iot-stats}.

\bibitem[Kristiyanto and Ernastuti, 2020]{deauth_analysis}
Kristiyanto, Y. and Ernastuti, E. (2020).
\newblock Analysis of deauthentication attack on ieee 802.11 connectivity based
  on iot technology using external penetration test.
\newblock {\em CommIT (Communication and Information Technology) Journal},
  14(1):45--51.

\bibitem[MITRE, 2006a]{CWE-287}
MITRE (2006a).
\newblock Cwe-287: Improper authentication.
\newblock \url{https://cwe.mitre.org/data/definitions/287.html}.

\bibitem[MITRE, 2006b]{CWE-341}
MITRE (2006b).
\newblock Cwe-341: Predictable from observable state.
\newblock \url{https://cwe.mitre.org/data/definitions/341.html}.

\bibitem[MITRE, 2008]{CWE-693}
MITRE (2008).
\newblock Cwe-693: Protection mechanism failure.
\newblock \url{https://cwe.mitre.org/data/definitions/693.html}.

\bibitem[MITRE, 2020]{CWE-1240}
MITRE (2020).
\newblock Cwe-1240: Use of a cryptographic primitive with a risky
  implementation.
\newblock \url{https://cwe.mitre.org/data/definitions/1240.html}.

\bibitem[Nath and Nath, 2022]{NATHN2022107997}
Nath, R. and Nath, H.~V. (2022).
\newblock Critical analysis of the layered and systematic approaches for
  understanding iot security threats and challenges.
\newblock {\em Computers and Electrical Engineering}.

\bibitem[Schepers et~al., 2022]{deaut_Wi-Fi6_WPA3}
Schepers, D., Ranganathan, A., and Vanhoef, M. (2022).
\newblock On the robustness of wi-fi deauthentication countermeasures.
\newblock In {\em Proceedings of the 15th ACM Conference on Security and
  Privacy in Wireless and Mobile Networks}, WiSec '22, page 245–256, New
  York, NY, USA. Association for Computing Machinery.

\bibitem[Wireshark, 2023]{wireshark}
Wireshark (2023).
\newblock {Wireshark project}.
\newblock \url{https://www.wireshark.org/}.

\end{thebibliography}
\end{document}